**Triplet *p*-wave superconductivity with ABM state in epitaxial Bi/Ni bilayers**

G. J. Zhao[1, 2*], X. X. Gong[3*], J. C. He[3*], J. A. Gifford[2], H. X. Zhou[3], Y. Chen[3,5], X. F. Jin[3,5], C. L. Chien[4,6,7], and T. Y. Chen[1, 2]

[1]Institute for Quantum Science and Engineering and Department of Physics, Southern University of Science and Technology, Shenzhen 518055, China

[2]Department of Physics, Arizona State University, Tempe, AZ  85287, USA

[3]Department of Physics and State Key Laboratory of Surface Physics, Fudan University, Shanghai 200433, China

[4]Department of Physics and Astronomy, Johns Hopkins University, Baltimore, MD 21218, USA

[5]Collaborative Innovation Center of Advanced Microstructures, Nanjing 210093, China

[6]Institute of Physics, Academia Sinica, Taipei, 11519, Taiwan

[7]Department of Physics, National Taiwan University, Taipei, 10617, Taiwan

* Authors have equal contributions for this work

Abstract

We report observation of spin triplet superconductivity in epitaxial Bi/Ni bilayers with $T_C$ up to 4 K and $2\Delta/k_B T_C \approx 12$. Andreev reflection spectroscopy (ARS) with ballistic injection of unpolarized and spin-polarized electrons conclusively reveals spin triplet $p$-wave superconductivity. The gap structure measured by ARS in multiple crystal directions shows the ABM (Anderson-Brinkman-Morel) state, the same as that in superfluid $^3$He.

Significance

Triplet $p$-wave superconductors, fundamentally important to understand unconventional superconductivity, are also the simplest systems to realize Majorana-zero-mode based quantum computing. Verified triplet superconductors are very rare, with $p$-wave pairing in $^3$He at 2.7 mK the only established example. Here we report the observation of $p$-wave triplet ABM (Anderson-Brinkman-Morel) state up to 4 K in epitaxial Bi/Ni bilayers, using Andreev reflection spectroscopy (ARS) with unpolarized and highly spin-polarized currents. ARS experiments in multiple directions determine the 3D gap structure and symmetry. The realization of $p$-wave superconductors in the solid state has important implication for topological Majorana physics and Weyl superconductivity.

Most superconductors (SCs) with known pairing states are singlet SCs [1], where the Cooper pairs have antiparallel spins, most commonly *s*-wave (e.g., Nb) with an isotropic energy gap ($2\Delta$), and some *d*-wave (e.g., high-$T_c$ cuprates) with an anisotropic gap structure with nodes [2]. Conspicuously lacking is triplet *p*-wave SCs, which feature prominently in realizing Majorana bound-state quantum computing [3, 4]. Some (e.g., $Sr_2RuO_4$) have been suspected to be, but remain unconfirmed as, *p*-wave SCs [5-7]. Superfluid $^3$He has been demonstrated as triplet *p*-wave pairing [8].

With the recent advent of topological superconductivity [9-11], the novel topological surface state may offer new routes to the elusive *p*-wave SCs. Since strong spin orbit coupling is an essential ingredient we focus our attention on materials that contain heavy elements, especially Bi, the heaviest non-radioactive element. Indeed, we have recently observed half quantum flux in $Bi_2Pd$, consistent with *p*-wave pairing [12]. In this work, we report the realization of triplet *p*-wave in epitaxial Bi(110)/Ni(100) bilayers.

To ascertain a triplet SC, thermodynamic measurements, e.g., upper crucial field, are suggestive but insufficient. NMR Knight shift to measure spin susceptibility across $T_c$ is often used [13]. Muon spin relaxation ($\mu$SR) [14] and magneto-optic Kerr effect (MOKE) [15] can detect time reversal symmetry (TRS) breaking across $T_c$, a necessary signature of triplet pairing. To ascertain the nature of a SC, one needs to measure the spin state of the Cooper pairs and the gap structure as Andreev reflection spectroscopy (ARS) can provide. We show in this work, ARS with ballistic injections of *both* unpolarized and highly spin-polarized electrons can reveal a singlet or triplet SC, and when administered in *multiple* crystalline directions, can also determine the 3D gap structure.

The defining difference between a singlet SC and a triplet SC is the spin states in the Cooper pair. The triplet *p*-wave pairing has three possible spin states, $|\uparrow\uparrow>$ ($S_z$ = 1), $|\uparrow\downarrow>+|\downarrow\uparrow>$ ($S_z$ = 0), and $|\downarrow\downarrow>$ ($S_z$ = -1). Triplet *p*-wave pairing has been established in superfluid $^3$He, in which there are A-phase and B-phase. In the B-phase, generally associated with the Balian-Werthamer (BW) state [16], the three spin states are populated with equal probability, resulting in an isotropic gap. For the A-phase associated with the Anderson-Brinkman-Morel (ABM) state [17, 18], the $S_z$ = 0 state is completely depleted. The ABM state, which can carry spin angular momentum and is related to quantum computing, is more interesting. For a large spin polarization of Cooper pairs, only one component $S_z$ = +1 exists with a supercurrent carrying a spin angular momentum. The gap of the ABM state is anisotropic as described by,

$$\hat{\Delta}(\vec{k}) = \Delta \begin{pmatrix} -e^{i\phi_k}\sin\theta_k & 0 \\ 0 & e^{i\phi_k}\sin\theta_k \end{pmatrix} \tag{1}$$

$$E_k = \left[ \varepsilon^2\left(\vec{k}\right) + \Delta^2 \sin^2\theta_k \right]^{1/2} \tag{2}$$

where $\theta_k$ and $\phi_k$ are spherical coordinates to describe the gap function. The 3D donut shape gap structure $\Delta \sin\theta_k$ with an extremal gap value $\Delta$ specifies the ABM state. The $\pm \phi_k$ halves have opposite signs. Previously, the ABM state has only been observed in $^3$He at temperature near 2.7 mK [4].

ARS has been utilized to determine spin polarization (*P*) of magnetic materials [19, 20], including some highly spin-polarized material [21, 22] and half metals [23, 24]. ARS has also been utilized to measure the gap of many s-wave SCs, MgB$_2$ with two gaps, and the Fe-SCs [25]. The ARS involves ballistic injection of electrons through a sharp tip from a metal into a SC or vice versa. When electrons are injected into a SC, only two electrons of appropriate spin can enter the gap as a Cooper pair. A normal metal with unpolarized electrons ($P = 0$, e.g., Cu and Au) has equal spin-up and spin-down bands at the Fermi energy $E_F$ [Fig. 1(a)]. Injection of a spin-up electron with $|E| < \Delta$ into a singlet SC must be accompanied by a spin-down electron, thus a reflected spin-down hole, the Andreev reflection process [26], and doubling the conductance within the gap (Fig. 1c). The same scenario occurs for a triplet SC as shown in Fig. 1. Thus, the ARS results for ballistic spin injection from a normal metal into either a singlet or a triplet SC are qualitatively the same with *increased* conductance within the gap as shown in Fig. 1(c). On the other hand, in a half metal with fully polarized electrons [$P = 1$, e.g., $CrO_2$ and $La_{2/3}Sr_{1/3}MnO_3$ (LSMO)], there is only one spin band [Fig. 1(b)] with no electron of opposite spin available at $E_F$, thus the Andreev reflection process is blocked with *depressed* conductance within the gap (Fig. 1d). However, in a triplet SC with parallel pairing, the Andreev reflection process is *allowed* for both a half-metal (Fig. 1b) *and* a normal metal (Fig. 1a) with *increased* conductance within the gap in both cases. Hence the AR spectra with a half metal, or a highly polarized metal, are qualitatively different for singlet and triplet SCs, suitable for clear identification.

The ARS results shown in Fig. 1 are for an ideal interface at $T = 0$ K with one-dimensional spin injection. Real contacts encounter interfacial scattering (Z), inelastic scattering ($\Gamma$), thermal smearing ($T$), extra resistance ($r_E$), and injection from all angles, which distort the ARS spectra and require more detailed analyses [27]. The ARS of an s-wave SC shows the same double-peak spectrum with injection from all angles, where the separation of the two peaks gives the gap value of $2\Delta$ [28]. For the ARS of *d*-wave or *p*-wave SCs with anisotropic gap structure, the ARS spectrum, the gap value and its phase depend on injection directions due to interference. The ARS spectra can be quantitatively fitted with the calculated the spectra for *s*-, *p*- and *d*-wave SCs as described in the Supplementary to obtain the gap value for each injection direction.

We use Au as the normal-metal tip and single crystal LSMO with *P* over 80% [29] as the half-metal tip.   For comparison, we first show results of *s*-wave Pb and *d*-wave YBCO in Fig. 2.  In both Pb and YBCO, the Au contact shows enhanced ARS spectra, whereas the LSMO contact shows depressed ARS spectra, revealing that both are singlet SCs.  The solid curves are quantitative fits and the obtained essentially the same gap values using Au or LSMO tip for the same SC.  The ARS results provide unequivocal identification of singlet SCs, either *s*-wave in Pb or *d*-wave in YBCO.   We next discuss the results on epitaxial Bi/Ni bilayers.

Co-existence of superconductivity and magnetism has previously been observed in polycrystalline Bi/Ni bilayers [30, 31].  The epitaxial Bi/Ni bilayers have been fabricated by molecular beam epitaxy (MBE) on either MgO (001) or Si(111) substrates to acquire respectively epitaxial Bi(110) or Bi(111) layers with various thicknesses of fcc Ni ($t_{Ni}$ = 0 - 7.5 nm) and rhombohedral Bi ($t_{Bi}$ = 0 - 500nm) [32].  The appearance of superconductivity depends sensitively on the epitaxial orientation.  Only Bi(110) exhibits superconductivity whereas Bi(111) does not.  For the Bi(110)/Ni(100) films, for the same thicknesses, the layer order of Bi/Ni and Ni/Bi is not important.  In the following, we discuss the results of Bi(110)/Ni(100).  Neither Bi nor ferromagnetic Ni, but remarkably Bi(20nm)/Ni(2nm), is superconducting with $T_C \approx$ 4 K as shown in Fig. S1(a).  Furthermore, Bi/Ni has a very unusual thickness dependence as shown in Fig. S1(d).  For a given Bi thickness, increasing Ni thickness always depresses $T_C$.  However, a thicker Bi layer readily restores the depressed $T_C$.  Thus, $T_C$ of every Bi thickness has its own dependence on Ni thickness.  In addition, the parallel upper critical field ($H_{C2\parallel}$) shows values exceeding the Pauli limit of 1.83 $T_C \approx$ 7 T as shown in Fig. S1(c), highly suggestive of an unconventional SC.

We use ARS with both Au and LSMO tips on the top Bi(110) surface of the Bi(20nm)/Ni(2nm) bilayer. As shown in Fig. 2(c), the ARS using *both* Au and LSMO show *enhanced* conductance within the gap, a defining characteristic of a triplet SC as discussed in Fig. 1. The ARS spectra can be well described quantitatively by the ABM state of triplet pairing, as shown by the solid curves. From the measured gap value of $\Delta \cong$ 2.2meV and $T_C$ = 4.0 K, we found $2\Delta/k_B T_C \approx$ 12, far larger than the value of $2\Delta/k_B T_C \approx$ 3.53, universally observed in BCS *s*-wave SCs [1], another unusual feature.

We measured many contacts with various contact resistances in the ballistic limit as shown in Fig. 3. When the Au tip is contact on the top surface (Fig. 3a), all the AR spectra with contact resistance from a few Ω to 1700 Ω show enhanced conductance (Fig 3b). Importantly, the AR spectra with a LSMO tip from a few Ω to over 1000 Ω also show enhanced conductance (Fig. 3d). We further show the measured $dI/dV$ of one Au contact (Fig. 3c) and one LSMO contact (Fig. 3e) at various temperatures. The peak structure of the enhanced conductance in both Au and LSMO contact vary systematically and disappears at $T_C$. At $T < T_C$, the normal state conductance ($dI/dV$)$_{normal}$ outside the gap remains finite for both Au and LSMO contacts, but decreases sharply near $T_C$. Concurrently, the peak intensity in both Au and LSMO contacts steadily reduce and vanishes at $T_C$, hence all originated from a triplet SC. The non-monotonous feature in the LSMO contact (Fig. 3e) may be due to the misalignment of the magnetizations of Ni and LSMO since it disappears in a small field of 1 kOe. These ARS results using injection from both Au and LSMO using many contacts and at various temperatures all indicate Bi/Ni is a triplet SC.

To determine the 3D gap structure, in addition to injection from the top surface, we measured ARS in the two other perpendicular directions of the epitaxial Bi/Ni samples cut along perpendicular crystal directions. Since point contact is unfeasible, we used Au or LSMO sharp blades to make contacts with Bi/Ni on the side of the sample, as schematically shown in Fig. 3f and 3i. The Ni magnetization was aligned by a magnetic field of 1 kOe and two directions of Fig. 3f and 3i are defined by the magnetization direction indicated by the arrow. The ARS spectra of the Bi/Ni bilayer are very different in the three directions, as summarized in Fig. 4(a-c). The spectra are similar in the A and C directions, but different in the B direction. This anisotropic gap structure rules out the triplet $p$-wave Balian-Werthamer (BW) state with a quasi-isotropic gap. The triplet $p$-wave ABM gap structure from Eq. (1) is an anisotropic 3D donut with its nodal direction along the B direction, as shown in the center of Fig. 4. In both A and C directions, the incident and reflected wavevectors have opposite $x$-component, thus an opposite phase from Eq. (1), giving rise to a single peak, as shown in Fig. 4(d), which is a cross section of the donut-like structure with two halves of opposite signs. One notes that the tunneling spectrum with a zero-bias coherent peak for $d$-wave SC (Fig. 2b) is due to the same physics [33]. In the B direction, however, the incident and reflected wave-vectors have the same sign, causing a double peak feature, as shown in Fig. 4(e). All the experimentally observed AR spectra can be excellently reproduced as the solid curves shown in Fig. 4(a,-c) using the ABM state.

To further establish the ABM structure, we measure the ARS at one direction (labeled B) using an in-plane magnetic field of 3 kOe (less than the upper critical field) to change the Ni magnetization to different direction. The ARS results with the Ni magnetization at angles of $\alpha = 0°, 45°, 90°, 135°, 180°$ are shown in Fig. 4(f). The results shows a single peak at $\alpha = 0°$, 180° (direction C), a double peak at $\alpha = 90°$ (direction B) and a triple peak at $\alpha = 45°$, 135°, which is the superposition of the previous two. These results conclusively show the two-fold axis of the ABM donut structure and that the Ni magnetization dictates its orientation. All the results in Fig, 4(f) have been observed from one contact, and all the determined gap values are consistently at $\Delta \cong 2.1$ meV. Recent Kerr measurement on Bi/Ni bilayers, showing spontaneous time-reversal symmetry breaking below $T_C$ [15], is consistent with the triplet nature. Our ARS measurements quantitatively demonstrate epitaxial Bi/Ni as the first material in which the triplet $p$-wave ABM state has been realized in the solid state.

The mechanism of the superconductivity in Bi/Ni, likely the results of topological superconductivity, is intimately related to both Bi and Ni. Bi is an unusual semimetal with bulk SC recently discovered at 0.53 mK [34]. The presence of ferromagnetic Ni initiates superconductivity with $T_C$ up to 4 K and the Ni magnetization direction dictates the orientation of the triplet $p$-wave ABM state. The itinerant Ni appears to be essential, since neither Bi/Co nor Bi/Fe show superconductivity. The ABM state has two gapless nodes on the Fermi surface which correspond to topological superconductor with two chiral edge Weyl points protected by Chern number as "magnetic" monopole [35]. As opposed to the ABM state in superfluid $^3$He occurring at less than 3 mK, many exploration and manipulation techniques can be readily utilized in the solid state of Bi/Ni at 4 K. The 3D triplet $p$-wave superconducting Bi/Ni bilayers may offer an ideal platform to explore topological Weyl superconductivity and Majorana physics.

This work was supported as part of SHINES, an EFRC center funded by the U. S. Department of Energy, Office of Science, Basic Energy Science, under award SC0012670.

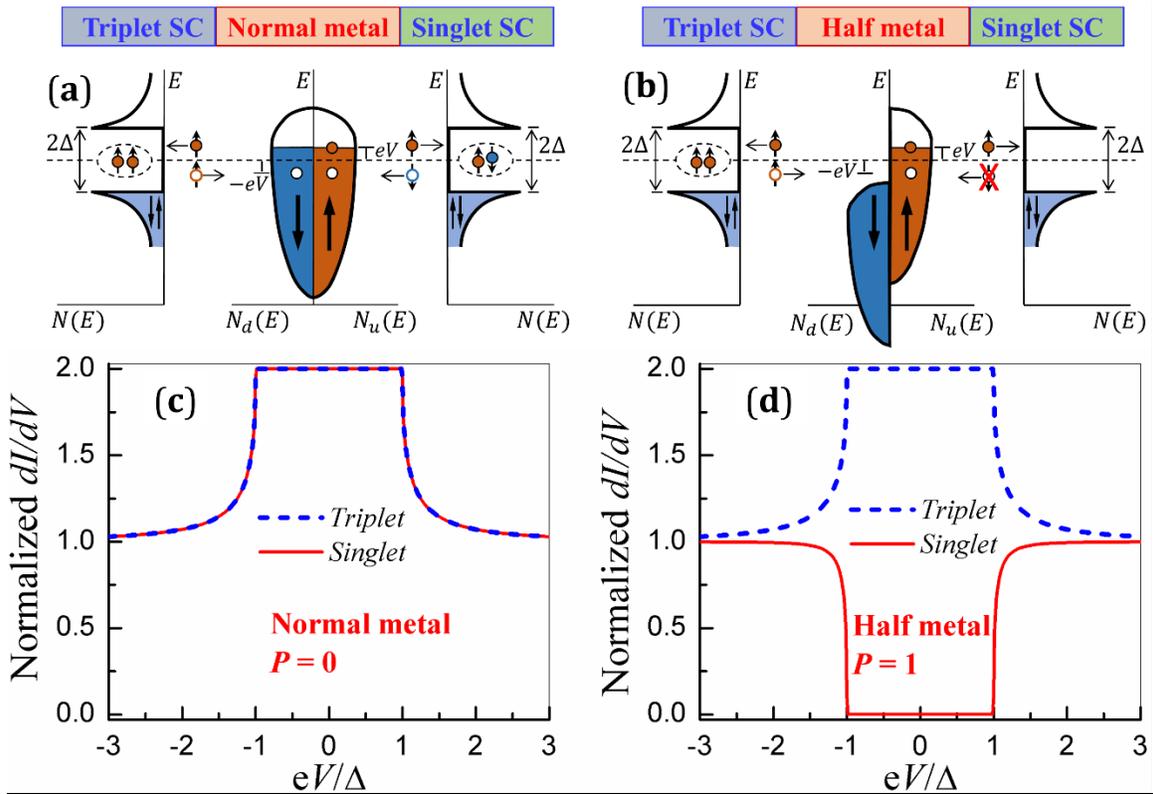

Fig. 1 Andreev reflection of a singlet and a triplet SC at $T = 0$ K using a normal metal ($P = 0$) and a half metal ($P = 1$) in 1D. Schematic energy diagrams of Andreev reflection of a singlet SC (right) and a triplet SC (left) using (a) a normal metal and (b) a half metal. Andreev spectra of a (c) singlet SC and (d) triplet SC, using a normal metal (dashed blue lines) and a half metal (solid red line) tip.

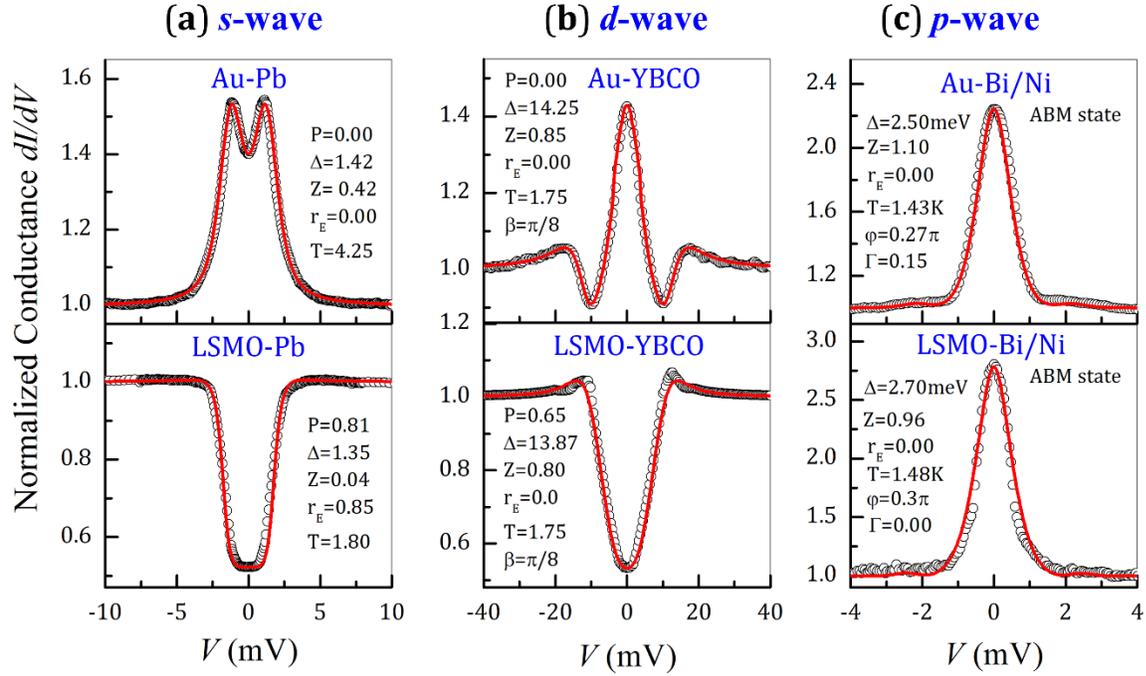

Fig. 2 Andreev spectra of (a) *s*-wave SC Pb, (b) *d*-wave SC YBCO tips, (c) *p*-wave SC of Bi/Ni fitted with the ABM state using Au and LSMO tips with fitted results as the solid curves.

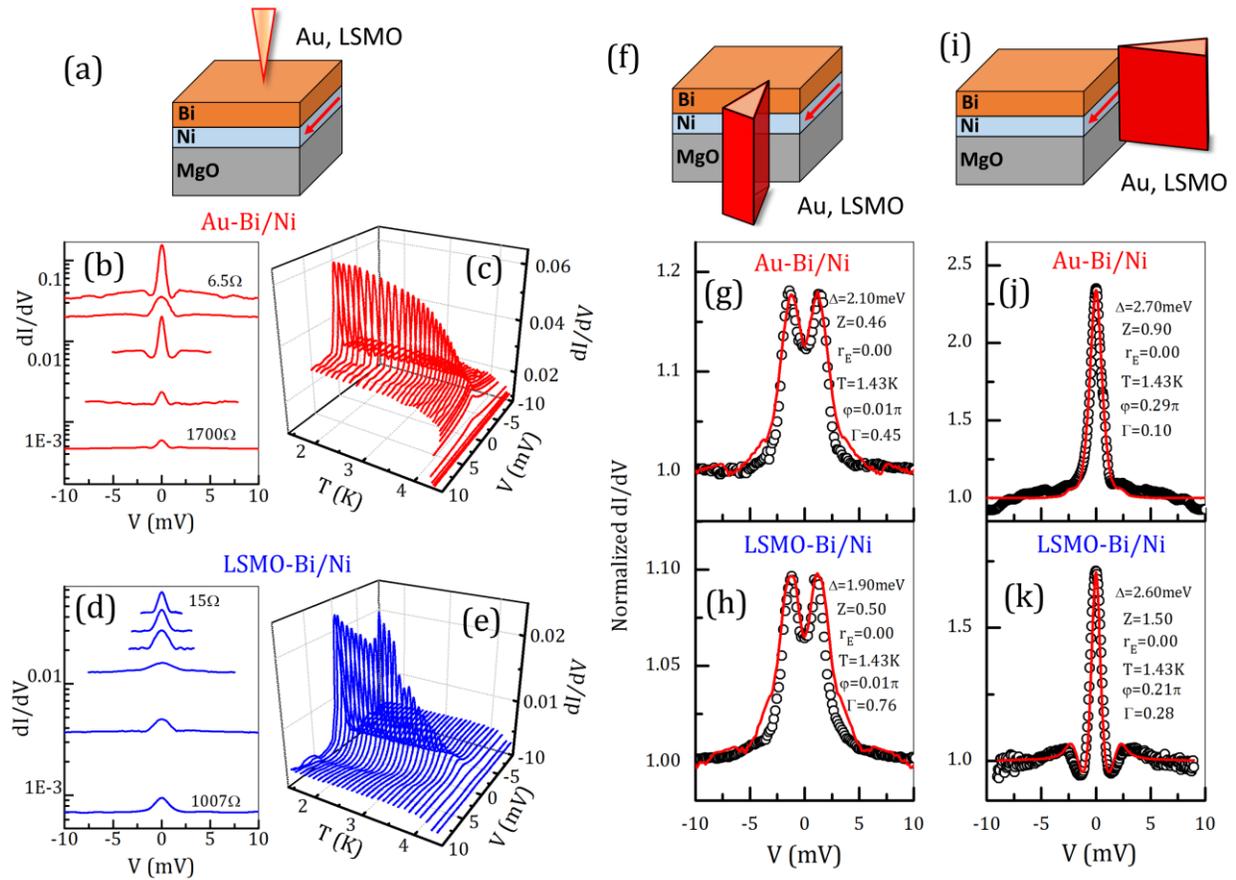

Fig. 3 ARS of Bi/Ni at 1.5 K with injection into the top surface (a-e), side surface (f-h) and other side surface (i-k). Results of injection from top surface using Au tip (red) with (b) various contact resistances, (c) one Au contact at various temperatures, and LSMO tip (blue) with (d) various contact resistances, (f) one LSMO contact at various temperatures. Representative ARS results on the side of Bi/Ni in the direction along the magnetization (**M**) direction of Ni with (g) Au tip and (h) LSMO tip. Representative ARS results on the side of Bi/Ni in the direction perpendicular to **M** with (j) Au tip and (k) LSMO tip. The remnant magnetization of the Ni layer is indicated by the red arrow.

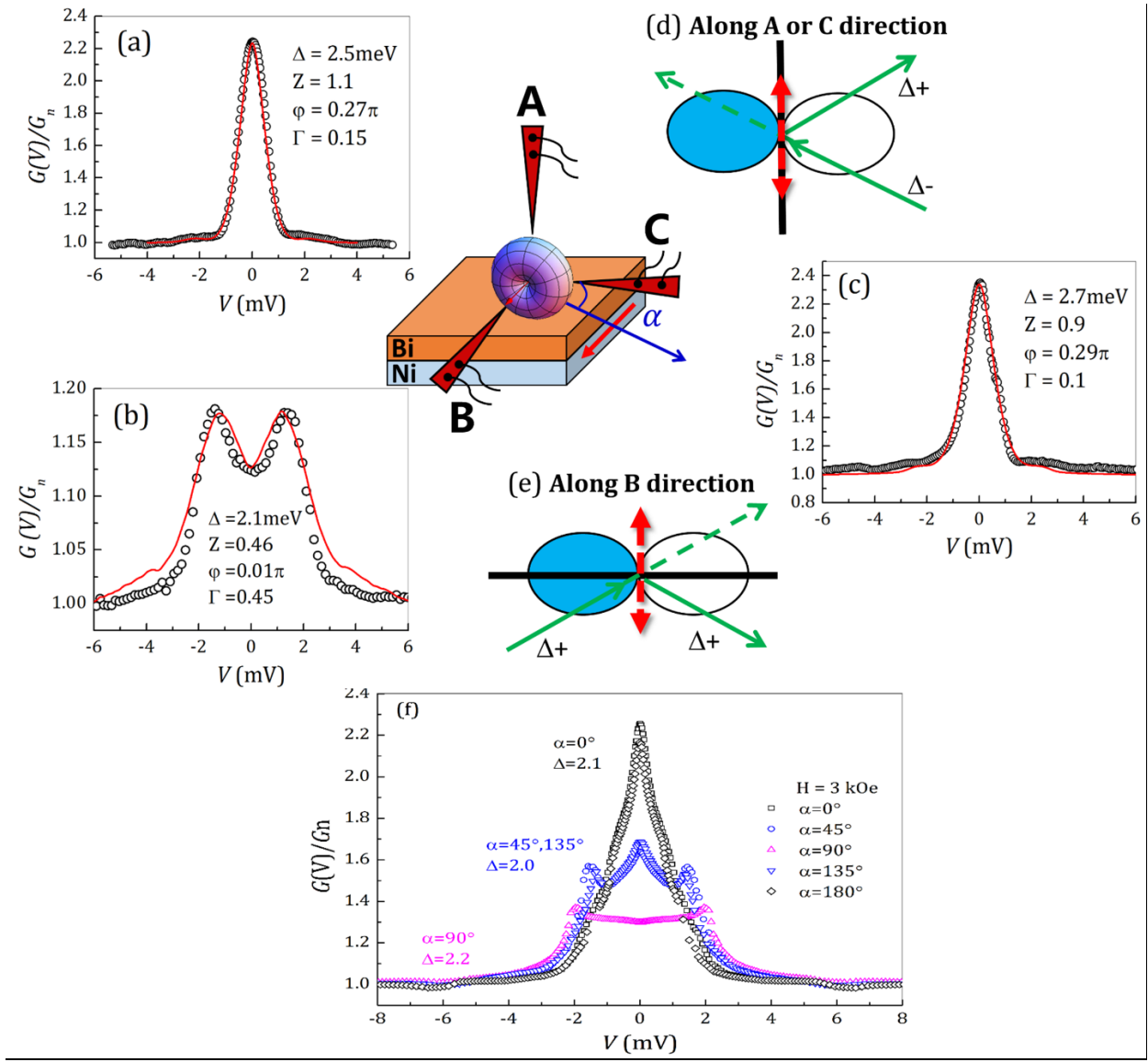

Fig. 4 Donut-like 3D ABM gap structure of the Bi/Ni bilayer with $E_F$ set to 0. (a-c), Andreev reflection spectra observed at the three directions A, B and C respectively. (d) Andreev reflection spectra at the A and C interfaces, where incident and reflection have opposite phase. (e) Andreev reflection spectra at B interface where incident and reflection have the same phase. (f) Andreev spectra of one point contact along **B** with an in-plane magnetic field of $H$ = 3 kOe ($H << H_C$) applied in different directions at angle $\alpha$ angle with respect to C with gap values indicated.

**Supplementary Materials**

**More Experimental Details**

The superconductivity in epitaxial Bi/Ni depends systematically on the layer thickness of both Bi and Ni, as summarized in Fig. S1(d). For a fixed thickness of Bi, e.g., $t_{Bi}$ = 20 nm, a Ni layer with increasing thickness always suppresses superconductivity. For $t_{Ni}$ = 2nm, $T_C$ is about 4 K but reduced to 2 K for $t_{Ni}$ = 4nm. Remarkably, increasing $t_{Bi}$ can *restore* the suppressed superconductivity and $T_C$ recovers back to about 4 K. This unusual thickness dependence cannot be simply explained by the assumption that the superconductivity occurs at the interface then decays into the Bi and Ni layers. In contrast, no superconductivity has been detected in either Bi/Co or Bi/Fe, illustrating the unique role of Ni in the system.

We have also measured the upper critical fields in Bi(20nm)/Ni(2nm) for both perpendicular ($H_{C2}^{\perp}$) (Fig. S1a) and parallel ($H_{C2}^{//}$) (Fig. S1b) to the film plane. To describe the field dependence, we use three characteristic temperatures of onset (0.95 $R_N$), mid-point (0.5 $R_N$) and zero-resistance (0.05 $R_N$), where $R_N$ is the normal state resistance. The $T$-dependences of the upper critical fields in the two directions are displayed in Fig. S1(c). Fitting the results of perpendicular field (open symbols) to the Werthamer–Helfand–Hohenberg formula of $H_{C2}^{\perp}(0) = 0.69 \mathrm{d} H_{C2}^{\perp} / \mathrm{d} T_{C} |_{T_C}$ [S1], the upper perpendicular critical field at $T = 0$ $K$ is about 1.9 T. For the parallel field (solid symbols), the upper critical field can be well described by $H_{C2}^{//}(T) = H_{C2}^{//}(0)(1 - T/T_{C})^{\alpha}$. The onset data (solid purple triangles) in Fig. S1C can be well described by exponent $\alpha$ = *2/3* (solid purple curve), as found in other quasi-2D SCs [S2,S3]. The upper parallel critical fields extrapolated to $T = 0$ $K$ are substantially above the Pauli limit of $B_{Pauli}$ = *1.83T$_C$* = 7.1T [S4-S6], which has often been taken as an indication of unconventional, including *p*-wave superconductivity. These features of Bi/Ni are highly unusual but conclusive triplet pairing requires spin injection within the gap.

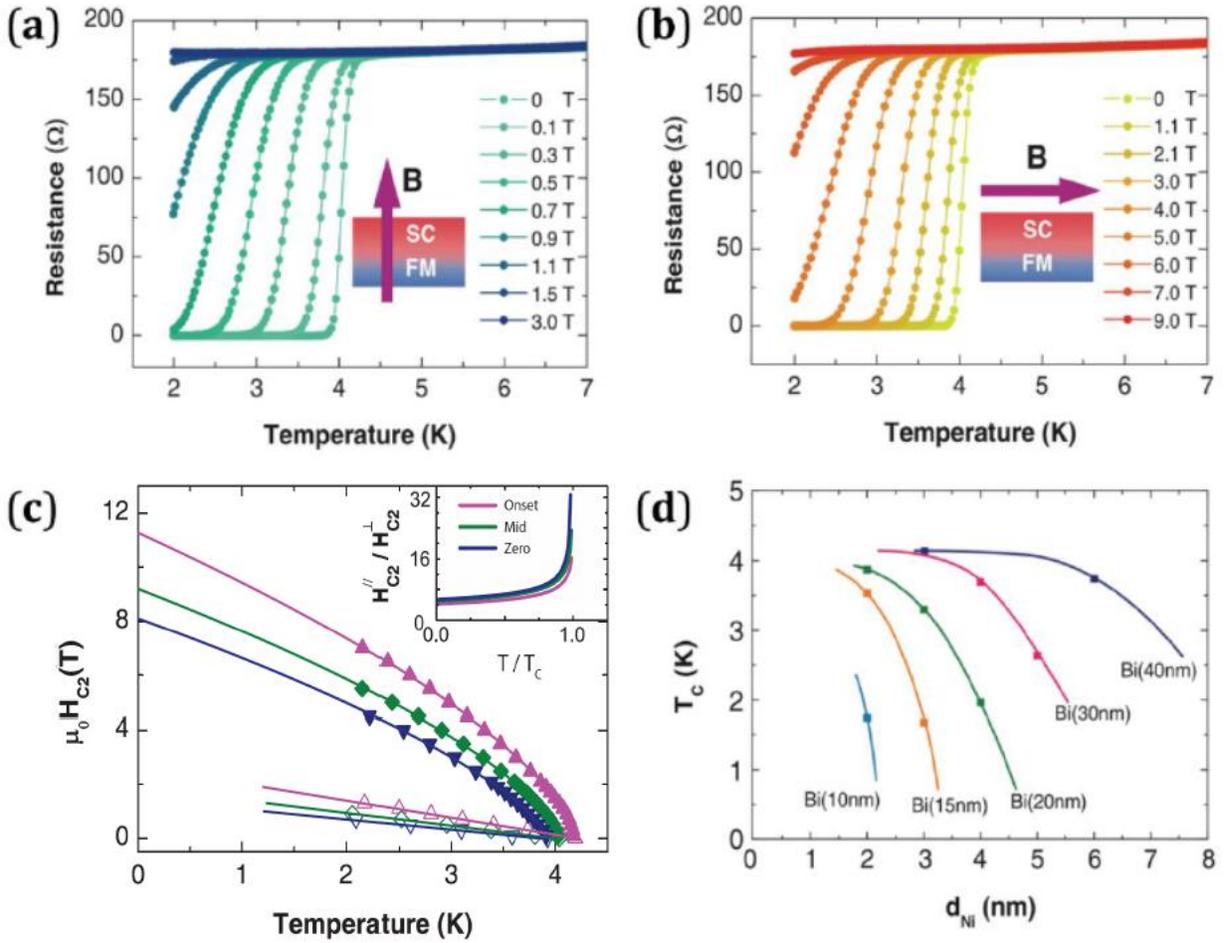

Fig. S1 | Superconducting properties of epitaxial Bi(110)/Ni(001)/MgO(001). Superconducting transition of Bi(20nm)/Ni(2nm)/MgO in (a) perpendicular and (b) parallel magnetic field. (c) Temperature dependence of upper critical fields $H_{C2}$ for perpendicular (open symbols) and parallel (solid symbols). (d) Dependence on superconducting transition temperature $T_C$ on Ni thickness for different Bi thicknesses. The lines are guides to the eyes.

**Theoretical Formalism**

As shown in Fig. 2(a) and (b), a spin polarized current from LSMO seriously suppresses the conductance in both s-wave and d-wave superconductors. But the conductance for the Ni/Bi bilayer is not suppressed but much larger than 2. This cannot be described by the singlet BTK model. Here we develop a triplet model to describe our data. For the Normal metal-Superconductor (N-S) junction [S7], the plane wave at normal metal side can be expressed by a four-component wave function $\psi_N(\vec{r})$,

$$\psi_N(\vec{r}) = e^{i\vec{k}_\parallel \cdot \vec{r}_\parallel} \begin{pmatrix} e^{ik_+x} + be^{-ik_+x} \\ 0 \\ a_2 e^{ik_-x} \\ a_1 e^{(\alpha+i)k_-x} \end{pmatrix} \tag{S1}$$

where $k_\parallel$ and $k_\pm$ are wave vector components parallel and vertical to interface respectively, the dimensionless real number $\alpha$ is related to spin polarization $P$, i.e. $P = \dfrac{\alpha^2}{4+\alpha^2}$. The first row of right hand side of Eq. (S1) represents the electron with spin up incident plane wave and normal reflection wave $e^{ik_+x} + be^{-ik_+x}$, the second row represents the electron with spin down wave $0$, the third row represents the hole with spin down Andreev reflection wave $a_2 e^{ik_-x}$ and the fourth row represents the hole with spin up evanescent wave $a_1 e^{(\alpha+i)k_-x}$. We can always choose the spin direction of incident electron as the positive $s_Z$ direction, the superconducting state could either be singlet pairing or triplet pairing, the coefficients ($a_1$, $a_2$ and $b$) can be solved from the boundary conditions below.

The wave function at the superconductor side is given by,

$$\psi_S(\vec{r}) = e^{i\vec{k}_\parallel \cdot \vec{r}_\parallel} \left\{ c e^{iq_+x} \begin{pmatrix} u_{\uparrow\uparrow}^+ \\ u_{\uparrow\downarrow}^+ \\ v_{\uparrow\uparrow}^+ \\ v_{\uparrow\downarrow}^+ \end{pmatrix} + d e^{-iq_-x} \begin{pmatrix} u_{\uparrow\uparrow}^- \\ u_{\uparrow\downarrow}^- \\ v_{\uparrow\uparrow}^- \\ v_{\uparrow\downarrow}^- \end{pmatrix} \right\} \tag{S2}$$

where $+$ and $-$ denote electron-like and hole-like respectively, $u_{\uparrow\uparrow}$, $u_{\uparrow\downarrow}$, $v_{\uparrow\uparrow}$, and $v_{\uparrow\downarrow}$ are coherence factors which come from the equation

$$a_{ks} = \sum_{s'} \left( u_{kss'} \alpha_{ks'} + v_{kss'} \alpha_{-ks'}^+ \right) \tag{S3}$$

where $s$ labels the spin index , $a_{ks}$, denotes annihilation operator of electron, $\alpha_{ks'}$ corresponds to annihilation operator of quasiparticle.

For unitary solution [S8] of a superconductor,

$$\hat{u}_k = \frac{[E_k + \epsilon(\vec{k})]\hat{\sigma}_0}{\left\{ [E_k + \epsilon(\vec{k})]^2 + \frac{1}{2} tr \hat{\Delta} \hat{\Delta}^+(\vec{k}) \right\}^{1/2}}$$

$$\hat{v}_k = \frac{-\hat{\Delta}(\vec{k})}{\left\{ [E_k + \epsilon(\vec{k})]^2 + \frac{1}{2} tr \hat{\Delta} \hat{\Delta}^+(\vec{k}) \right\}^{1/2}} \tag{S4}$$

where $E_k = \left[ \epsilon^2(\vec{k}) + \frac{1}{2} tr \hat{\Delta} \hat{\Delta}^+(\vec{k}) \right]^{1/2}$, and $\hat{\Delta}(\vec{k})$ is a $2 \times 2$ pairing matrix. For singlet pairing case

$$\hat{\Delta}(\vec{k}) = i\hat{\sigma}_y \psi(\vec{k}) = \begin{pmatrix} 0 & \psi(\vec{k}) \\ -\psi(\vec{k}) & 0 \end{pmatrix} \tag{S5}$$

where $\psi(\vec{k})$ is an even function. For triplet pairing case

$$\begin{aligned} \hat{\Delta}(\vec{k}) &= i(\vec{d}(\vec{k}) \cdot \hat{\sigma})\hat{\sigma}_y \\ &= \begin{pmatrix} -d_x(\vec{k}) + i d_y(\vec{k}) & d_z(\vec{k}) \\ d_z(\vec{k}) & d_x(\vec{k}) + i d_y(\vec{k}) \end{pmatrix} \end{aligned} \tag{S6}$$

where $\vec{d}(\vec{k})$ is an odd vector function.

The boundary conditions satisfy

$$\psi(x = 0^+) = \psi(x = 0^-),$$

$$\frac{\hbar^2}{2m} \psi'_x \Big|_{x=0^-} - \frac{\hbar^2}{2m} \psi'_x \Big|_{x=0^+} + U\psi \Big|_{x=0} = 0. \tag{S7}$$

For ballistic transport, the normalized conductance with a bias voltage $V$ is

$$\sigma(eV) = \frac{\int_{-\pi/2}^{\pi/2} g^T(eV) \cos\theta \, d\theta}{\int_{-\pi/2}^{\pi/2} g^T(\infty) \cos\theta \, d\theta} \tag{S8}$$

where $g^T(eV) = \int_0^1 g\left(|eV + \frac{1}{\beta} ln\frac{1-f}{f}|\right) df$, $g(E) = 1 + |a_1(E)|^2 + |a_2(E)|^2 - |b(E)|^2$ and $\beta = 1/k_B T$. Considering the effects of additional resistance on Andreev reflection spectrum, we adopted a self-consistent method for extracting $dI_{NS}/dV_{AR}$ from the measured $dI_{NS}/dV$ [S9].

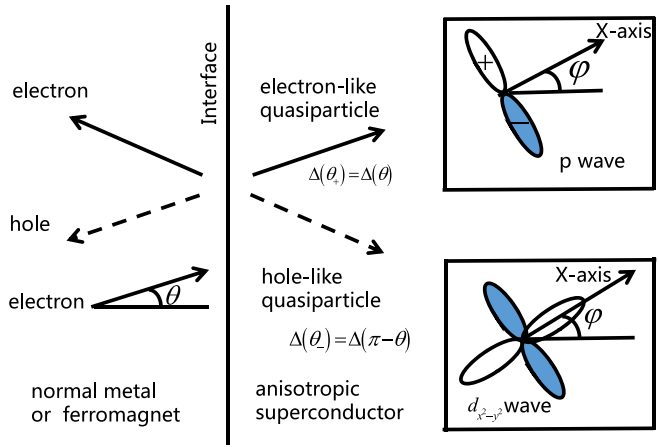

FIG.S2: The parameter $\theta$ depicts the electron incident angle and $\varphi$ represents the angle between the $x$ axis of the p-wave (or $d_{x^2-y^2}$-wave) and the normal direction of interface. The transmitted hole-like quasiparticle and electron-like quasiparticle have different effective pairing potentials $\Delta(\theta_+)$ and $\Delta(\theta_-)$, respectively, with $\theta_+ = \theta$ and $\theta_- = \pi - \theta$.

As studied in the superfluid state of ³He, there are two thermodynamically stable spin triplet p-wave pairing states [S10]. One is the Anderson-Brinkman-Morel (ABM) state [S11] with its superconducting order parameter as

$$\hat{\Delta}(\vec{k}) = \Delta \begin{pmatrix} -e^{i\phi_k}\sin\theta_k & 0 \\ 0 & e^{i\phi_k}\sin\theta_k \end{pmatrix} \tag{S9}$$

We can easily obtain $\hat{\Delta}(\vec{k})\hat{\Delta}^+(\vec{k}) = \Delta^2\sin^2\theta_k\hat{\sigma}_0$ which imply the solution of ABM state is unitary. So we can impose Eq. (S7) to obtain its coherence factors which are elements of matrices $\hat{u}_k = \sqrt{\frac{1}{2}(1+\epsilon(\vec{k})/E_k)}\hat{\sigma}_0$, $\hat{v}_k = sign(\sin\theta_k)e^{i\phi_k}\sqrt{\frac{1}{2}(1-\epsilon(\vec{k})/E_k)}\hat{\sigma}_z$ $(u_{\uparrow\downarrow}^{\pm} = v_{\uparrow\downarrow}^{\pm} = 0)$. The obtained coefficients $a_1$, $a_2$, $b$, $c$, $d$ are listed in Table S. The normalized conductance spectra of ABM state are presented in Fig. 2 and Fig. 3. The solid red lines are the best fit using our model. Our theoretical analysis of experimental data illustrates that there is ABM state in Bi/Ni bilayer system.

As shown in Fig.4, the 3D gap of ABM state is like a donut and there are two nodal points. Meanwhile, its nodal direction is parallel to the B direction, which is same as the Ni magnetization direction. We find the fitting value of superconducting gap in B direction is slightly smaller than that of A and C directions (the fitting values of superconducting gap in A and C directions are quite close). We think that experimentally the point contact in B direction may change the local orientation of the surface so that the nodal direction of ABM state may deviate slightly from the normal direction of the surface. This may cause the pair breaking effect [S12] and result in a smaller superconducting gap in B direction.

The Balian-Werthamer (BW) state [S13], the so-called B phase in ³He, is another typical candidate of p-wave pairing states. Owing to the isotropy of the single particle spectrum, the BW state is in many respects similar to the s-wave pairing state. Our theoretical results show that BW state is unable to explain the main features of Andreev reflection experimental results.

The ABM state, or the A-phase, is an equal spin pairing state, where only Cooper pairs with $S_z=+1$ and $S_z=-1$ exist, while the Sz=0 component is absent. The external magnetic field may imbalance the population of $S_z=+1$ and $S_z=-1$ Cooper pairs and result in the emergence of the so-called $A_2$-phase with finite spin polarization of Cooper pairs. If the spin polarization of Cooper pairs becomes quite large, only one component $S_z=+1$ exists, this state is called the $A_1$-phase. So if one progressively increases the external magnetic field, there may be two phase transitions from A-phase, to $A_2$-phase and then to $A_1$-phase. It is rather interesting that all the gap parameters of A, $A_1$ and $A_2$ phases have the same axial structure or nodal structure, which reflects the topological nature of these pairing states[S10].

In Bi/Ni bilayer, Ni layer acts not only as a pairing inducer due to enhanced ferromagnetic spin fluctuation in Bi layer but also a pairing breaker due to the static Ni ferromagnetism. Because of the Ni layer, in principle all three topological phases (A, $A_1$ or $A_2$) could be the candidate. As shown in Fig. 3 and Fig. 4, in the absence of external magnetic field, the nice data fitting based on A-phase may suggest that the ground state is close to A-phase. The formation of 20nm thick non-superconducting Bi layer sandwiched by Ni layer and Bi superconducting layer may greatly suppress the external magnetic field exerted by Ni layer to Bi superconducting layer. That may explain why Bi superconducting layer corresponds to equal spin pairing state or ABM state. In the presence of 3 kOe external magnetic field, the Bi superconducting state may enter into $A_2$-phase with finite spin polarization of the Cooper pairs.

TABLE S: Coefficients for ABM state. $Z = mU/\hbar^2 k_F$ , $p = (Z^2 + 1)u^+_{q\uparrow\uparrow}v^-_{q\uparrow\uparrow} - Z^2 u^-_{q\uparrow\uparrow}v^+_{q\uparrow\uparrow}$ ,

$\hat{u}^{\pm}_q = \sqrt{\frac{1}{2}\left(1 \pm \epsilon^{\pm}(\vec{q})/(|E_q| + i\Gamma)\right)}\hat{\sigma}_0$ , $\hat{v}^{\pm}_q = sign(\sin\theta^{\pm}_q)\sqrt{\frac{1}{2}\left(1 \mp \epsilon^{\pm}(\vec{q})/(|E_q| + i\Gamma)\right)}\hat{\sigma}_z$ ,

$\epsilon^{\pm}(\vec{q}) = \sqrt{(|E_q| + i\Gamma)^2 - \Delta^2(\theta^{\pm}_q)}$ , $\Delta(\theta^+_q) = \Delta\sin(\theta_q - \varphi)$, $\Delta(\theta^-_q) = \Delta\sin(\pi - \theta_q - \varphi)$, $\theta^+_q = \theta - \varphi$,

$\theta^-_q = \pi - \theta - \varphi$ (illustrated in Fig.S2) and $\theta_q \in (0, 2\pi)$, $\Gamma$ representing inelastic electron-electron scattering factor.

| | $a_1$ | $a_2$ | $b$ | $c$ | $d$ |
|---|---|---|---|---|---|
| *ABM state* | 0 | $\dfrac{v^+_{q\uparrow\uparrow}v^-_{q\uparrow\uparrow}}{p}$ | $\dfrac{Z(Z+i)(u^-_{q\uparrow\uparrow}v^+_{q\uparrow\uparrow} - u^+_{q\uparrow\uparrow}v^-_{q\uparrow\uparrow})}{p}$ | $\dfrac{-i(Z+i)v^-_{q\uparrow\uparrow}}{p}$ | $\dfrac{iZv^+_{q\uparrow\uparrow}}{p}$ |

**Some possible superconducting Bi-Ni materials**

One might suspect the triplet superconductivity in Bi/Ni bilayers is due to superconducting Bi-Ni alloys or compounds at the Bi/Ni interface. The NiBi$_3$ is indeed superconducting with $T_c$ of about 4 K, but decidedly *s*-wave [S14].